\documentstyle[12pt]{article}
\def\lsim{<\kern-2.5ex\lower0.85ex\hbox{$\sim$}\ }
\def\rsim{>\kern-2.5ex\lower0.85ex\hbox{$\sim$}\ }    
\def\LAMBDABAR {\hbox{$\lambda$\kern-0.52em\raise+0.45ex\hbox{--}\kern+0.2em}}

\begin{document}
\rightline{UR-1529\ \ \ \ \ \ \ \ }
\rightline{ER/40685/917}
\ \ \ \ \ 

\vskip 1truein

\begin{center}
{\bf ALTERNATIVES TO AXION PRODUCTION \\
\smallskip
 AND DETECTION$^*$}
\end{center}
\bigskip
 \bigskip
\begin{center}
{by}
\end{center}
\bigskip
\bigskip
\begin{center}
{A.C. Melissinos\\
\smallskip
Department of Physics and Astronomy \\
\smallskip
 University of Rochester \\ 
\smallskip
 Rochester, NY 14627, USA}
\end{center}

\vskip 3truein

\noindent $^*$To appear in Proceedings of the 5th IFT Workshop on Axions, 
University of
Florida, Gainesville, FL, March 13-15, 1998.

\vfill\eject

\noindent{\bf 1. Heavy Ion Collisions}

\bigskip

Axions are associated with the Peccei-Quinn symmetry which is broken at 
very high energies $(f \rsim 10^{10}$ GeV).
 However axions acquire mass near
 the QCD phase transition $( E \sim 1$ GeV). Thus it should be possible to 
produce axions in heavy ion collisions (at RHIC and the LHC) where the 
equivalent \lq\lq temperature" of the quark-gluon plasma will exceed the 1 
GeV mark. We estimate the expected rate of axion production and detection 
as follows \cite{sikivie}.

The number density of axions at time $t_1$, where $T(t_1) = T_1 \sim 1$ GeV 
is given by
\begin{eqnarray}
n_a(t_1) &\simeq& \rho_a (t_1)
 {1 \over m_a (t_1)} \nonumber \\
\noalign{\vskip 4pt}%
&\simeq& {1 \over 2} v^2 
m_a (t_1) < \alpha^2 (t_1)> \nonumber \\
\noalign{\vskip 4pt}%
&\simeq&  \pi f^2_a {1 \over t_1} = {1 \over 2} f^2_a m_a  
\end{eqnarray}
For the variables used in Eq. (1) we note that $v$ and $\alpha$ are related 
to the vacuum expectation value of the scalar field $\phi (x)$

$$<\phi (x)>\  = v e^{i \alpha (x)}$$

\noindent and thus
 the axion field is given by

$$a(x) = v \alpha (x)$$

\noindent We have also introduced
 the axion coupling $f_a~=~v/N$ where $N$ is the \lq
\lq number of minima" in the axion effective potential, and 
$<~\alpha~(x)>~ 
\ \sim~1/N$. Furthermore

$$f_a m_a \simeq f_\pi m_\pi$$

\noindent where $f_\pi$, $m_\pi$ are
 the pion decay constant and mass, for $\sim 0.1$ GeV,
$m_\pi \sim 0.14$ GeV. Finally we can 
take

$$t_1 m_a (t_1) \simeq 2 \pi$$

\noindent Thus we expect  the axion
 density in a quark-gluon plasma at $T \sim
 1$ GeV to be $(1\ ~{\rm eV}\ ~=~5~\times~10^4\ ~{\rm
cm}^{-1})$
\begin{eqnarray}
n_a & = & {1 \over 2} (f_\pi m_\pi)^2
 {1 \over m_a} \simeq 10^{37} \bigg(
{10^{-5}\ {\rm eV} \over m_a} \bigg) \ {\rm eV}^3 \nonumber \\
\noalign{\vskip 4pt}%
&\simeq& 10^{12} \bigg( {10^{-5}\ {\rm 
eV} \over m_a}\bigg)\ {\rm F}^{-3}
\end{eqnarray}

To detect the axions produced we will use
 their conversion to photons in a 
magnetic field \cite{ruoso}. The conversion probability is given by 
\begin{equation}
P_{a \rightarrow \gamma}
 = {1 \over 4} g_{a \gamma \gamma}^{2} (B 
\ell)^2
\end{equation}
with $B$, $\ell$ the magnetic
 field strength and length and $g_{a \gamma \gamma}$ 
the axion to 2-photon coupling
\begin{equation}
g_{a \gamma \gamma} =
 {\alpha \over 2 \pi} {N \over f_a}
\end{equation}
Combining Eqs. (1,3 and 4) we find the probability for detection of a 
photon from axion conversion
\begin{equation}
P_\gamma = \big( {\alpha
 \over 2 \pi}\big)^2 {N^2 \over 8} (B \ell)^2
m_a V
\end{equation}
where $V$ is the volume of the quark gluon plasma region. It is interesting 
that Eq. (5) depends on the axion field parameters linearly and only
through the axion mass $m_a$.

We evaluate Eq. (5) for typical parameters $B~=~ 5$~ T =~ $10^3$~ eV$^2$,

$$\ell = 10\ {\rm  m}\  = 5 \times 10^7\ {\rm eV}^{-1},$$

$$V \simeq (10 \ {\rm F})^3 = 10^{-22}\ {\rm  eV}^{-3}$$

\noindent and set

$$(\alpha /2\pi)^2 (N^2/8) \simeq 10^{-6}$$

\noindent to find

\begin{equation}
P_\gamma = 2.5 \times 10^{-7} \bigg({m_a \over
 1 {\rm eV}}
\bigg)\big/{\rm collision}
\end{equation}

\noindent Assuming single bunches in
 RHIC and one central collision per beam crossing 
$(5 \times 10^4$ collisions/second) the expected photon rate is

\begin{equation}
P_\gamma \sim 0.7 \bigg( {\textstyle m_a \over 1 {\rm eV}}
\bigg)/{\rm minute}
\end{equation}

In writing Eq. (5) we assumed that all axions produced during the quark-
gluon formation period cross the magnet and reach the detector.  Such a 
condition can be approached by exploiting the Lorentz contraction for 
asymmetric collisions. If $A$ is the area and $L$ the distance of the 
detector from the collision point, we must have

\begin{equation}
1 / \gamma^2_{\rm cm} \geq A/L^2
\end{equation}

\noindent For typical values
 $A \sim 100$ cm$^2$, $L= 10$ m one finds $\gamma > 100$.  
This is easy to achieve in a fixed target experiment but in a collider it 
will be necessary to optimize the trade-off between cm energy and detector 
acceptance.

Finally we estimate the energy of the detected photons. Presumably in the 
cm system the axion energy will be representative of the plasma 
temperature, $E_a \sim 1$ GeV.
 If the collision is asymmetric the 
 additional boost
results in higher energies at the detector.
 Thus one is searching for high energy photons (multi-GeV) 
which appear to have traversed through the shielding surrounding the 
collision point.  Such a signal is unambiguous, especially since it depends 
quadratically on the value of the converting magnetic field.

\bigskip

\noindent {\bf 2. Critical Field}

\bigskip

It has been recently shown that in a critical electromagnetic field,
 $e^+ e^-$ pairs are spontaneously produced \cite{burke,acm}.
 The critical field is 
defined \cite{schwinger} by

\begin{equation}
E_c = {m^2 c^3 \over e\hbar} = 1.3 \times 10^{16}\ 
{\rm V/cm}
\end{equation}
and diagrammatically, the pair production can be viewed as the multiphoton 
process shown in Fig.~1a. The pair can

\vfill\eject

\ \ \ \ 

\vskip 2.5 truein

\noindent couple to produce an axion as shown in Fig. 1b, if the $e \hbox{-}
e \hbox{-}a$ coupling is different from zero.

The probability per unit time and unit volume for graph (a) 
and for strong fields is given by \cite{brezin}

\begin{equation}
w_{e^+ e^-} = {\alpha E^2 \over \pi \hbar^2} \exp \bigg( - 
{\pi m^2 c^3 \over e \hbar E}\bigg)
\end{equation}
For graph (b) we multiply Eq. (10) by the  $e \hbox{-}
e \hbox{-}a$ coupling squared and by an inverse area typical of the 
interaction region. Since the $e^+ e^-$ pair is produced at one space-time 
point within a volume of typical dimensions of the electron's Compton 
wavelength, we write
\begin{equation}
w_a = {\alpha E^2 \over \pi \hbar^2} 
{m^2_e c^2 \over \hbar^2} \bigg( {\hbar c \over f_a}\bigg)^2
\exp \bigg( - 
{\pi m^2 c^3 \over e \hbar E}\bigg)
\end{equation}
For $E \simeq \pi E_c$ the number of axions produced in volume-time 
$VT$ is 
given by
\begin{equation}
N_a \simeq \bigg( {m_e \over f_a}\bigg)^2 {Vc T \over 
\LAMBDABAR^4_c}
\end{equation}

Near-critical fields in the laboratory frame should be reached by focusing 
to the diffraction limit the beam from the next generation of 1${\rm \AA}$ 
X-ray Free Electron Lasers \cite{free}. 
 We then take $V = (10{\rm \AA})^3$ and $T = 0.2$ ps with a 
repetition frequency of $7~\times~10^4$~Hz (for the DESY FEL), to obtain an 
axion rate
\begin{equation}
R_a = 5 \times 10^{-9} \bigg( {m_a \over {\rm eV}}\bigg)^2\ {\rm s}^{-1}
\end{equation}
This is too low a rate for considering laboratory experiments since the 
axion detection efficiency is equally small. 

Axion production in a critical field could, however, be important in 
astrophysical settings where the magnetic field of neutron stars can reach 
critical value. If this is the case, axions with an energy $E_a \sim 1$ MeV 
should be emitted.  If we assume that $V$ in Eq. (12) is given by the entire 
volume of the neutron star $(R = 10$ km) located at a distance of 0.3 kpc, 
then we obtain for the flux on earth
\begin{equation}
F_a \simeq 5 \bigg( {m_a \over {\rm eV}}\bigg)^2
\ {\rm  m^{-2}s^{-1}}
\end{equation}
Still an excessively low rate, but the process could be of interest if a 
proper calculation reveals the possibility of higher fluxes.

\vfill\eject

\end{document}